\documentclass[12pt]{iopart}

\renewcommand{\>}{\rangle}
\newcommand{\<}{\langle}

\usepackage{graphicx}

\newcommand{\ba}{\begin{align}}
\newcommand{\ea}{\end{align}}
\newcommand{\la}{\label}
\newcommand{\be}{\begin{equation}}
\newcommand{\ee}{\end{equation}}
\newcommand{\bea}{\begin{eqnarray}}
\newcommand{\eea}{\end{eqnarray}}
\def\12{\frac{1}{2}}
\newcommand{\p}{\partial}

\renewcommand{\phi}{\varphi}

\begin{document}

\title[Quantum hydrodynamics and nonlinear differential equations \ldots]{Quantum hydrodynamics and nonlinear differential equations for degenerate Fermi gas}

\author{Eldad Bettelheim}
\address{Racah Institute of Physics, The Hebrew University of
Jerusalem, Safra Campus, Givat Ram, Jerusalem, Israel 91904.}
\ead{eldadb@phys.huji.ac.il}

\author{Alexander G. Abanov
}
\address{Department of Physics and Astronomy,
Stony Brook University,  Stony Brook, NY 11794-3800.}
\ead{alexandre.abanov@sunysb.edu}

\author{Paul B. Wiegmann}
\address{James Franck Institute
of the University of Chicago, 
5640 S.Ellis Avenue, Chicago, IL 60637.}
\ead{wiegmann@uchicago.edu}

\begin{abstract}
We present new nonlinear differential equations for spacetime correlation functions of Fermi gas in one spatial dimension. The correlation functions we consider describe non-stationary processes  out of equilibrium. The equations we obtain are integrable  equations.  They generalize  known nonlinear differential equations for  correlation functions at equilibrium \cite{McCoy, 1980-Perk, Miwa, Korepin} and provide  vital tools to study non-equilibrium dynamics of electronic systems. The method we developed  is based only on Wick's theorem and the hydrodynamic description of the Fermi gas. Differential equations appear directly in bilinear form.
\end{abstract}
\date{\today}

\maketitle

\section{Introduction}
Various spacetime correlation functions of free Fermi gas in one spatial dimension (and models  related to it -- impenetrable Bose and Fermi gases, $XY$-spin chain, 2D-Ising model, etc.) are Fredholm determinants or their minors and, therefore, generally speaking, obey integrable nonlinear differential  equations  with respect to spacetime and other parameters like temperature, chemical potential, etc. In practice, a derivation   of these equations is a complicated task.  Few equations are known. However, once obtained, they are indispensable in studies of off-shell properties of Fermi gas.

Historically nonlinear differential  equations for correlation functions first appeared in studies of the 2D Ising model \cite{McCoy}.  In Ref.  \cite{1980-Perk} these equations has been extended, applied to the XY-spin chain, and have been   derived in explicitly bilinear form with the help of Wick theorem generalized for these purposes in Ref. \cite {1977-PerkCapel}.  In Refs. \cite{Miwa}  
it  has been  showed that at zero temperature the two-point, equal-time  correlation function of impenetrable bosons and of the $XY$-spin chain can be expressed through Painleve transcendents. These equations have been extended to time- and temperature-dependent correlation functions at equilibrium for impenetrable bosons and the $XY$-chain  in Refs. \cite{Korepin,Korepin2}.   
 
The derivation of  the known   equations was essentially based on equilibrium and translational invariance properties of correlation functions. Although they have been recognized as integrable, their relations to integrable hierarchies remained unclear.

Recently, interest focused on electronic systems out of equilibrium \cite{LevitovAbanin} and especially  on non-stationary properties of  propagating localized states  \cite{paperfour}.
There the physical properties  change drastically and the dynamics acquires essentially  nonlinear features such as hydrodynamic instabilities \cite{paperIII,paperfour}. Apart from being of a fundamental interest, research in non-equilibrium degenerate Fermi gas is driven by a quest  to transmit quantum information in electronic nanodevices and to control and manipulate entangled quantum many-body states.  Integrable  equations describing the non-equilibrium -- non-stationary states of  electronic gas, once available, would be a valuable tool  in studies of a complex dynamics of electronic gas. They are derived in this letter.

We will obtain nonlinear differential equations with respect to spacetime for the correlation functions of vertex operators
\be
 \la{vertex}
	e^{ia_L\phi_L}\cdot e^{ia_R\phi_R},
\ee
where  $\phi_{R,L}$ are right/left handed chiral Bose fields of fermionic currents and $a_{L,R}$ are  arbitrary (not necessarily real) parameters.  The definition of these fields in terms of free fermions is given below. Equations with respect to parameters of the density matrix (e.g., temperature or chemical potential) can be also obtained, but are not discussed here.

The correlation functions of vertex operators (\ref{vertex}) are important in many applications. In the case of equilibrium they also satisfy some previously known equations. A few particular cases are worth mentioning. The case $a_L=a_R=1/2$ describes a one-point function of impenetrable bosons \cite{Miwa,Korepin}. The imaginary $a_L=a_R$ and $a_L=-a_R$ yield a generating function of moments of the distribution function of  the density and velocity of electronic gas.  The chiral correlation functions  $a_L=0$ or $a_R=0$ describe a  tunneling into an electron gas (Fermi Edge Singularity) \cite{FES}, etc.

In this article we give a simple {\it physical} derivation of equations on these correlations functions. Our derivation is based solely on a general form of the Wick theorem  \cite{comment21}, the representation of Fermi-operators through Bose fields (see e.g., \cite{stone}), and the hydrodynamic description of Fermi gas \cite{Riemann}. Our equations appear directly in Hirota's bilinear form \cite{comment22}. They are known as {\it modified Kadomtsev-Petviashvili} (or mKP) equations and are the first and the second equation of mKP hierarchy. In a particular case, in free fermion systems these equations have  previously appeared in \cite{paperI,paperIII}.  The nonlinear equations, being integrable are proven to be an effective tool in computing asymptotes of  correlation functions  in different regimes. These computations are specific  to each particular problem and will not be discussed  here (see, Refs. \cite{paperIII,paperfour} for some practical  applications).

In the following after necessary preliminaries on the Fermi-Bose correspondence, we describe the correlation functions (\ref{7}),  present the nonlinear differential equations they obey (\ref{Hirota},\ref{Hirota3},\ref{Hirota4}), and then give a short derivation of those equations.

\section{Fermions and a Bose Field}
We consider free fermions on a circle  of a circumference $L$
\begin{equation}\label{H}
H= \sum_p\frac{p^2}{2m}\psi_p^\dag\psi_p,
\end{equation}
where $\psi_p$ 
is a  mode of a fermion
field $\psi(x)=\frac{1}{\sqrt{ L}}\sum_p  e^{\frac{i}{\hbar}px}\psi_p $.

A central point of our approach is the hydrodynamic description of the Fermi gas. We briefly review it.
The properties of  the Fermi gas  are fully described in terms of  canonical hydrodynamic variables: density and velocity
\bea
	\rho=\frac{1}{2\pi}\left(J^R+J^L\right), \quad  {v}=\frac{ \hbar}{2m}\left( J^L-J^R\right),
  \nonumber
\eea
where the chiral (right, left) currents are
\bea
	J^{R,L}(x)=\frac{2\pi}{ L}\sum_k e^{\frac{i}{\hbar}kx}J_k^{R,L},\quad
	J_k^{R,L}=\sum_{\pm p> 0} \psi_{p}^\dag\psi_{p+k}.
 \nonumber
\eea
Tomonaga's  equal time commutation relations of currents (see, e.g., \cite{stone})
\bea
	[J^R_k,J^R_{l}]=[J^L_k,J^L_{l}]= \frac{Lk}{2\pi\hbar}\delta_{k+l,0},\quad [J^R_k,J^L_l]=0
 \la{currentalgebra}
\eea
lead to a canonical relation between density and velocity:
\bea
	[\rho(x,t),\, {v}(y,t)]=-i \frac{\hbar}{m}\nabla\delta(x-y)
 \label{algebra}
\eea
and to the canonical Bose field
\bea
 \nonumber
	\nabla \phi_{R,L} \equiv \pm J^{R,L}(x,t)=\frac{m}{\hbar}{v} \pm \pi \rho    , \quad [\phi_L,\phi_R]=0.
\eea
We will be interested  in the spacetime dependence of correlations of two vertex operators:
\begin{equation}
 \label{7}
	\tr\Big(\!\!\!:\!e^{-ib_L\phi_Lt(\xi')-ib_R\phi_R(\xi')}
	\!: :\!e^{ia_L\phi_L(\xi)+i a_R\phi_R(\xi)}\!:\!\!\varrho\Big).
\end{equation}
Here $\varrho$ is the density matrix and $\xi\!\!=\!\!(x,t)$.  The colon
denotes  normal ordering with respect to the vacuum \cite{comment1}.

Below we drop the chiral subscripts $R,L$
in all formulas and assume that the upper (lower) sign corresponds to the right (left) sectors.

\section{Coherent states}
We will consider correlation functions with respect to {\it coherent states}. This means that the density matrix is an element of $gl(\infty)$, i.e., it is an exponent bilinear in fermionic modes:
\bea
 \nonumber
	\varrho=\exp\left({\sum_{p,q} A_{pq}\psi^\dag_p \psi_q}\right).
\eea
This choice of the density matrix allows us to use Wick's theorem.

The set of coherent states is  rather general. It  exhausts most interesting applications. For example, the density matrix can be  a  Boltzmann distribution, or any other non-equilibrium distribution as in \cite{LevitovAbanin}. Another choice would be $\varrho = |V\>\< V|$,  where $|V\>=\exp\left({\sum_{k} t_kJ_k}\right)|0\>$ is a coherent state (i.e., an eigenstate of current operators).
Such states appear as a result of  shake up of the electronic gas by electromagnetic potential with harmonics $t_k$.

\section{Quantum Hydrodynamics}
The Hamiltonian of free fermions (\ref{H}) in the sector of coherent states can be expressed  solely in  terms of density and velocity \cite{Riemann}
\begin{equation}\label{5}
H=\int\left(\frac{m\!:\!\!\rho v^2\!\!:}{2}+\frac{\pi^2\hbar^2}{6m}\!:\!\rho^3\!:\right) dx.
\end{equation}
Evolution of   density and velocity follows
\bea
 \nonumber
	 :\! \dot\rho\!:  + \nabla\!:\! ( \rho v ) \!: =0\quad\quad &&\mbox{Continuity equation} 
 \nonumber\\
	:\!\dot v\!:  + \frac{1}{2} \nabla\!:\! ( v^2 +\frac{\pi^{2}\hbar^{2}}{m^{2}}\rho^2 )\!:=0 \quad\quad &&
	\mbox{Euler's equation}
 \nonumber
\eea
The hydrodynamic equations
are decoupled into two independent chiral quantum Riemann equations
\begin{equation}
 \label{Riemann}
	\dot \phi  + \frac{\ \hbar}{2 m} \!:\!(\nabla \phi )^2 \!:=0 .
\end{equation}

\section{Vertex operator and correlation functions}
The Heisenberg evolution equations  for the chiral vertex operators follow from the quantum Riemann equation  (\ref{Riemann}) (see \ref{eqvertex})
\bea
 \label{1}
	 ( i\partial_t - \frac{\hbar}{2m} \nabla^2  ) \!:\!
	e^{ ia\phi}\!:
	&=& \frac{\hbar}{2m} a(a+1)\!:\! e^{ia     \phi} T\!:,
 \\
 \label{2}
	 (i\partial_t + \frac{\hbar}{2m} \nabla^2  ) \!:\!
	e^{ i(a+1) \phi}\!\!:
	&=& -\frac{\hbar}{2 m}a(a+1)\!:\!e^{ i(a+1) \phi}  \bar{T} \!:,
\eea
where $T=\, :\!(\nabla\phi)^2\!:\!-i\nabla^{2}\phi$ and  $\bar{T}=\,:\!(\nabla\phi)^2\!:\!+i\nabla^{2}\phi$ are holomorphic (antiholomorphic) components of the stress-energy tensor of a chiral Bose field (with the central charge $1/2$).

A particular consequence of  these equations is that
the vertex operators $e^{\pm i\phi_{R,L}}$ obey the  Schroedinger equation. This yields to a familiar formula representing fermions as exponents of a Bose field
\bea
 \label{F}
	\!:\!e^{i\phi}\!:\simeq \sqrt{L} \psi(x),\quad \!:\!e^{-i\phi}\!:\simeq \sqrt{ L} \psi^\dagger(x).
\eea
A useful representation for current $J=\pm  {:\!e^{-i\phi}(-i\nabla) e^{i\phi}\!:}$ and   the stress energy tensor
reads
\bea
 \label{T}
	T=-:\!e^{-i\phi}\nabla^2 e^{i\phi}\!:,\quad  \bar{T}=-:\!e^{i\phi}\nabla^2 e^{-i\phi}\!:
\eea
Current and  stress energy tensor  can be   further cast in fermionic form using (\ref{F}) and OPE for vertex operators as follows from the current algebra (\ref{algebra})\cite{comment0}
\bea
 \label{OPE}
	&\!:\!e^{ia\phi(x)}e^{\!-\!ib\phi(x')}\!:=\!\left(\!\frac{\mp i2\pi (x\!-\!x')}{L}\!\right)^{ab}\!\!\!\!\!:\!e^{ia\phi(x)}\!::\!e^{-ib\phi(x')}\!\!:
\eea
With the help of (\ref{OPE}, \ref{T}) and (\ref{F}) we write
\bea
 \label{T1}
	T=\mp 4\pi i \psi^\dagger \nabla \psi,
	\quad \bar{T} = \mp 4 \pi i \psi \nabla\psi^\dagger,
	\quad J = 2\pi \psi^\dagger \psi.
\eea

Finally we cast the equation of motion of the vertex operator in a suggestive  bilinear form (Quantum Hirota equation)
\begin{equation}
 \label{qHirota}
	\left(iD_t -  \frac{\hbar}{2m}D^2_x\right) \!:\!e^{ ia\phi}\cdot e^{i(a+1)\phi}\!\!:=0.
\end{equation}
Here $D$ is the Hirota derivative $D f \cdot g = \partial f
g - f \partial  g$ and $D^2 f \cdot g = \partial^2 f
g -2\p f\partial g+f\p^2 g $.
This equation follows from the   quantum Riemann equation (\ref{Riemann}).

\section{Nonlinear differential equations}
The nonlinear equations themselves do not depend on the parameters $a_{L,R},\,b_{L,R}$ in (\ref{7}).   For brevity  we put $L=2\pi$ in formulas below and consider only the chiral and neutral ($a=b$) correlation function, which we denote as
\begin{equation}
 \label{tau}
	\tau_{a}(\xi,\xi')=\tr\left( :\!e^{-ia\phi (\xi')}\!::\!e^{ia\phi(\xi)}\!:\varrho\right).
\end{equation}

Being presented in  bilinear form the differential equations  for correlation functions  look identical to the quantum equation  (\ref{qHirota}):
\bea
 \label{Hirota}
	\left( i D_t - \frac{\hbar}{2m} D_x^2 \right) \tau_{a} \cdot \tau_{a+1}=0.
\eea
The equations must be viewed as equations for $\tau_{a+n}(x,t; x',t')$ in the continuous variables $x,t$ and the discrete variable $n$, where $|a|\leq 1/2$. As such they are a closed set of equations known as the mKP equations.
The correlation function can be found given initial conditions at $t=0$, for all integer $n$ and any $x$. Certain asymptotes or analytical conditions may also be helpful in solving the problem for particular cases. This program is taken up in practice in  \cite{paperIII,paperfour}, where in \cite{paperIII} the density matrix  had the form $|A\>\<0|$ where $|0\>$ is the ground state of electron gas and $|A\>$ is  a chiral coherent state of the form $|A\> = \varrho|0\>$. We also comment that $\tau_a$ where $|a|\leq 1/2$  is a Fredholm determinant, while the $\tau_{a+n}$  are its minors.

The mKP equations (\ref{Hirota}) has many different forms discussed in the literature. We will mention one of them. Setting
\bea
 \nonumber
	\Phi = i \frac{\hbar}{m} \log\left(\frac{\tau_{a}}{\tau_{a+1}} \right), \qquad
	\;\tilde{\Phi} = \frac{ \hbar}{m}  \log (\tau_{a} \tau_{a+1})
\eea
the equation (\ref{Hirota}) becomes
\bea
 \label{BO}
	\partial_t \Phi=-\frac{1}{2} (\nabla \Phi)^2+ \frac{\hbar}{2m}\nabla^2 \tilde{\Phi}.
\eea
In this form the classical equation (\ref{BO}) resembles the quantum equation (\ref{Riemann}). The last term in (\ref{BO})  can be considered as a quantum correction to the semiclassical Riemann equations.

In addition to the evolution  (\ref{Hirota}) there is another equation which does not involves time evolution.  This equation connects correlation functions with different density matrices: $\varrho$ and
$\tilde\varrho=\sum_{pq}a_{pq}\psi^\dag_p\varrho\psi_q$, where $a_{pq}$ is an arbitrary non-degenerate matrix.
Then $\tau_a$ and $\tilde\tau_a=\tr\left( :\!e^{-ia\phi (\xi')}\!::\!e^{ia\phi(\xi)}\!:\tilde\varrho\right)$
are related by
\be
 \label{Hirota3}
	D_xD_{x'}\tilde \tau_a\cdot \tau_a
	=  a^2 (\tilde\tau_{a+1} \tau_{a-1}+ \tau_{a+1}\tilde \tau_{a-1}).
\ee
This equation is a minor generalization  of familiar 2D Toda lattice equation.

An important particular case occurs when the modes $p,q$ in $\tilde\varrho$ lie very far from the Fermi surface. Then $\tilde\tau_a\sim \tau_a$ and the equation (\ref{Hirota3})  becomes 1D Toda chain equation
\be
 \label{Hirota4}
	D_xD_{x'} \tau_a\cdot \tau_a = 2 a^2\tau_{a+1} \tau_{a-1}.
\ee
This equation, together with  (\ref{Hirota}) being written  for pairs $\tau_a, \tau_{a+1}$ and $\tau_{a-1},\tau_a$ give a closed set  which involves only $\tau_{a-1}, \tau_a, \tau_{a+1}$.

\section{Translation invariant states}
Further closure can be achieved for  special cases. For example, if the density matrix  commutes with momentum, the correlation function   depends on the difference  $x-x'$ (e.g., the density matrix depends only on occupation numbers $n_p=\psi^\dag_p\psi_p$, $\varrho=\exp{\sum_p\lambda_pn_p}$. Then it also commutes with the Hamiltonian, is a function of  $t-t'$, and describes stationary processes). In this case we denote the tau-function as
\bea
 \nonumber
	\tau_{a}(x,t;x',t') = F_{a} (x-x',t,t').
\eea
As follows from (\ref{Hirota},\ref{Hirota4}) three functions $F_{a-1},\, F_a, \,F_{a+1}$ obey  three equations with respect to $(x,t)$
\bea
	 \left( i D_t - \frac{\hbar}{2 m} D_x^2 \right) F_a\cdot F_{a+1} &=& 0,
 \nonumber\\
	 \left( i D_{t} + \frac{\hbar }{2 m} D_{x}^2 \right) F_{a}\cdot F_{a-1} &=& 0,
 \nonumber\\
	 D_x^2 F_a\cdot F_a = -2 a^2 F_{a-1} F_{a+1}. &&
 \label{HirotaF2}
\eea
Here we used $D_{x'}=-D_x$.

These equations are the bilinear form of the Nonlinear Schroedinger equation (without a complex involution).
Introducing  $\Psi=F_{a+1}/F_{a}$ and  $\bar\Psi=F_{a-1}/F_{a}$ we have
\bea
	\left(i\p_t+\frac{\hbar}{2m} \nabla^2\right)\Psi
	&=& \frac{\hbar}{m} a^2\tilde\Psi\Psi^2,
 \nonumber \\
	\left(-i\p_t+\frac{\hbar}{2m} \nabla^2\right)\tilde\Psi
	&=& \frac{\hbar}{m}a^2\Psi\tilde\Psi^2.
\eea
These equations were obtained for time and temperature dependent two-point correlation function for impenetrable bosons  at equilibrium ($\varrho=e^{-H/T}$) \cite{Korepin}. The impenetrable bosons are equivalent to free fermions. The creation operator of the boson is the vertex operator (\ref{vertex}) with $a_L=a_R=1/2$.

\section{Derivation of Nonlinear equations} 
The fact that the quantum equation for vertex operators (\ref{qHirota}), and the classical equations for the correlation functions, (\ref{Hirota}), have the same form may not be an accident but may reflect a general property of integrable equations written in bilinear form. Here we limit our discussion by demonstrating  this phenomena for the equation (\ref{Hirota}).

We prove now the main formula (\ref{Hirota}). First we adopt the notation:
\bea
	\langle \mathcal{O}_1 (x), \mathcal{O}_2(x') \rangle
	=\left( \tau_{a}(x,t;x',t')\right)^{-1}
 \nonumber \\
	\times \tr \left( \varrho\, e^{i H t} \!:\! e^{ia \phi(x)} \mathcal{O}_1(x) \!:\!
	e^{-i H (t-t')} \!\!:\!e^{-ia \phi(x')} \mathcal{O}_2(x')\!:\! e^{-i H t'}  \right),
 \nonumber
\eea
where only the spatial dependence has been specified in the r.h.s..

Let us multiply  (\ref{1}) and (\ref{2})  by $:\!e^{-ia\phi(x',t')}\!:$  and  $:\!e^{- i (a+1)\phi(x',t')}\!:$ respectively, take the trace, and substitute into Eq. (\ref{Hirota}). We find that (\ref{Hirota})  holds due to the identity
\bea
 \label{W}
	&  \< e^{-i\phi},e^{i\phi}\bar{T} \> + \< \textbf{1},T \> \<e^{-i\phi},e^{i\phi}\>
	=   2\< \textbf{1},J \> \<  e^{-i\phi},e^{i\phi}J \>,
\eea
where $\bf{1}$ is the identity operator, and we further suppressed the arguments $x,x'$.

To prove this identity we first  write $:\!e^{i\phi}  \bar T\!:$ in terms of fermions.  Using (\ref{T}) we write
$\!:\!e^{i\phi} \bar T\!:=-\lim_{y,z\to x}\nabla_y^2 \!:\!e^{i\phi(x)}e^{i\phi(z)}e^{-i\phi(y)}\!:$, where we split spatial points. We now use the formula (\ref{OPE}) for vertex operators
and Eq. (\ref{T}) to write
\begin{eqnarray}
	\!:\!e^{i\phi(x)}  \bar T(x)\!:
	&=& \mp i (2\pi)^{3/2}  
	\lim_{y,z\to x}\nabla_y^2 \frac{(y-z)(y-x)}{(z-x)} \psi(x)\psi(z)\psi^\dag(y).
\end{eqnarray}
This  formula  allows to write the first term in (\ref{W}) in terms of a correlator with  four fermions insertion
\begin{equation}
 \label{longone}
	\mp  i (2\pi)^{2} \lim_{y,z\to x}\nabla_y^2 \frac{(y-z)(y-x)}{(z-x)}
	\<    \psi^\dagger (x'), \psi(x) \psi(z) \psi^\dag(y) \>.
\end{equation}
Now we use a general form of the Wick theorem \cite{comment21} to express the four fermion correlator in terms of correlators containing two-fermions
\bea	
	\<   \psi^{\dagger}(x'), \psi(x) \psi(z) \psi^\dagger (y) \>
	&=& \<   \psi(x) ,  \psi^\dagger (x') \> \<    \psi^\dag(y) \psi(z) ,  {\bf 1} \>
 \nonumber \\
	&-& \<   \psi(z) ,  \psi^\dagger (x') \> \<    \psi^\dag(y) \psi(x) ,  {\bf 1} \>.
  \label{wicks}
\eea
After this, we trace back  the path which lead us to (\ref{longone}). With the help of Eqs.  (\ref{F}) and (\ref{OPE}) we obtain the formulas
\bea 
	\<   \psi^{\dagger}(x') ,  \psi(x) \>  
	  &=& \frac{1}{2\pi}\<  e^{-i\phi(x')} ,e^{i\phi(x)} \>, 
 \nonumber \\
	\<  \textbf{1},  \psi(z)\psi^\dag(y)  \>
	&=& \pm \frac{1}{2\pi i (y-z)} \<  \textbf{1}, e^{i\phi(z)} e^{-i\phi(y)}  \>
 \label{OPE1}.
\eea
We use them in order to write  the r.h.s. of  (\ref{wicks})  in terms of bosons.  After that, one applies the operation $\lim_{y,z\to x}\nabla_y^2 \frac{(y-z)(y-x)}{(z-x)} $ to the  r.h.s of (\ref{wicks}),  takes the derivative,  and merges the points. This leads  to  (\ref{W}). The calculations are somewhat cumbersome, but straightforward \cite{comment}.

We briefly comment on the proof  of (\ref{Hirota4}), the proof of the more general  (\ref{Hirota3}) goes along the same lines.  First we write (\ref{Hirota4}) in the form
\bea
 \label{HirotaF2explained}
	\<J,J\>  - \< J,\textbf{1} \> \<  \textbf{1} ,J\> =
	 \< e^{i\phi }, e^{-i\phi}\> \< e^{-i\phi } ,e^{i\phi }\>.
\eea
 The proof of this equation is easier: after replacing bosonic  exponentials and the current operator by fermionic operators according to (\ref{F},\ref{T1}) one recognizes  Wick's theorem.

\section{Integrable hierarchy  of nonlinear equations}
As a final remark we mention that higher equations of the mKP hierarchy  describe evolution of Fermi gas
with an arbitrary spectrum $H=\sum_p \varepsilon_p\psi_p^\dag\psi_p$. They are  generated by the Hirota equation described in \cite{paperI}. Equations for magnetic chains equivalent to the Fermi gas can be obtained in a similar manner.

\section{Acknowledgment}
PW thanks V. Korepin for discussion of the results of this paper. We thank J. H. H. Perk for pointing out to us and discussing the results of \cite{1980-Perk, 1977-PerkCapel}.
 EB was supported by ISF grant number 206/07.
The work of AGA was supported by the NSF under the grant DMR-0348358.
PW was supported by NSF under the grant  DMR-0540811 and MRSEC DMR-0213745.

\appendix

\section{Equation of motion for vertex operators}
\label{eqvertex}
In this Appendix we sketch the derivation of equations (\ref{1},\ref{2}) from the Riemann equation (\ref{Riemann}). First of all we show how to calculate the time derivative of the vertex operator. A subtlety  here is that $\dot{\phi}$ does not commute with $\phi$. 

We recall a simple consequence of  the Hadamard lemma. If $H$ and $ A$ are two operators such that
$\left[\left[H,A\right],A\right] $ commutes with $A$, then
\bea
	\left[H,e^{A}\right] &=& e^{A}\left(\left[H,A\right]
	+\frac{1}{2!}\left[\left[H,A\right],A\right] 
	\right) 
 	= \left(\left[H,A\right]
	-\frac{1}{2!}\left[\left[H,A\right],A\right] 
	\right)e^{A}.
 \nonumber
\eea
If $H$ is a Hamiltonian $\partial_{t}A\sim \left[H,A\right]$ and we obtain for the time derivative of an exponent
\bea\la{A2}
	\partial_{t}e^{A} &=& e^{A}\left(\dot{A}
	+\frac{1}{2!}\left[\dot{A},A\right] 
	\right) 
 	= \left(\dot{A}
	-\frac{1}{2!}\left[\dot{A},A\right] 
	\right)e^{A}.
\eea
Let us now compute an evolution of $e^{ia\phi}$. We assume that $\phi$ is a right chiral field.  The left chiral field obeys the identical equation. 

We notice that the commutator $\left[\left[\dot{\phi}^{+},\phi^{+}\right],\phi^{+}\right]$ vanishes, and then apply (\ref{A2}). We obtain
\bea
	\partial_{t}:e^{ia\phi}: &=& \left(\partial_{t}e^{ia\phi^{+}}\right) e^{ia\phi^{-}}
	+ e^{ia\phi^{+}}\left(\partial_{t}e^{ia\phi^{-}}\right)
 \nonumber \\
 	&=& e^{ia\phi^{+}}\left(ia\dot{\phi}^{+} 
	+\frac{(ia)^{2}}{2!}\left[\dot{\phi}^{+},\phi^{+}\right] 
	\right) e^{ia\phi^{-}}
 \nonumber \\
 	&+& e^{ia\phi^{+}}\left(ia\dot{\phi}^{-} 
	-\frac{(ia)^{2}}{2!}\left[\dot{\phi}^{-},\phi^{-}\right] 
	\right) e^{ia\phi^{-}}.
 \la{intermexpr}
\eea

Let us now compute the commutator $\left[\dot{\phi}^{+},\phi^{+}\right]$. Using (\ref{Riemann}) and (\ref{currentalgebra}) we obtain for Fourier components
\be
	\left[\dot{\phi}_{p},\phi_{q}\right] = \frac{1}{m}(p+q)\phi_{p+q}.
\ee
Then we proceed as
\bea
	\left[\dot{\phi}^{+}(x),\phi^{+}(x)\right] 
	&=& \frac{1}{m}\left(\frac{2\pi}{L}\right)^{2}\sum_{p,q<0}
	e^{\frac{i}{\hbar}(p+q)x}(p+q)\phi_{p+q}
 \nonumber \\
	&=& \frac{1}{m}\left(\frac{2\pi}{L}\right)^{2}\sum_{k<0}
	e^{\frac{i}{\hbar}kx}k\phi_{p}\sum_{k<p<0}1
 	= - \frac{1}{\hbar m}\frac{2\pi}{L}\sum_{k<0}e^{\frac{i}{\hbar}kx}k^{2}\phi_{k}
 \nonumber \\
	&=& \frac{\hbar}{m} \nabla^{2}\phi^{+}.
 \nonumber
\eea
Repeating the same calculation for $\phi^{-}$ we obtain
\be
	\left[\dot{\phi}^{\pm}(x),\phi^{\pm}(x)\right] = \pm \frac{\hbar}{m} \nabla^{2}\phi^{\pm}.
 \la{dotcommut}
\ee
Finally, using (\ref{dotcommut}) in (\ref{intermexpr}) we obtain  (\ref{1},\ref{2})
\bea
	\partial_{t}:e^{ia\phi}: &=& e^{ia\phi^{+}}\left(ia\dot{\phi} 
	+\frac{(ia)^{2}}{2}\frac{\hbar}{m}\nabla^{2}\phi
	\right) e^{ia\phi^{-}}
 \nonumber \\
	&=& \frac{\hbar}{2m} e^{ia\phi^{+}}\left(-ia:(\nabla\phi)^{2}:
	+(ia)^{2}\nabla^{2}\phi
	\right) e^{ia\phi^{-}}.
\eea




\begin{thebibliography}{99}
\bibitem{comment1}
	The normal ordering  of hydrodynamic modes reads: 
	the negative modes of the right chiral current $J^R_k,\,k<0$ and 
	positive modes of  left chiral current $J^L_k,\,k>0$ are placed to the left to the rest of the modes. 
	For example $:\!e^{ia\phi}\!:= e^{ia\phi_+}e^{ia\phi_-}$, where
	$\phi_-(x) =\mp i \frac{2\pi\hbar}{L} \sum_{\pm k>0} \frac{1}{k}J_k e^{\frac{i}{\hbar}kx}$ and 
	$\phi_+(x) = \mp i \frac{2\pi\hbar}{L} \sum_{\pm k<0} \frac{1}{k}J_k e^{\frac{i}{\hbar}kx}$.
\bibitem{comment0} 
	In this and further formulas we assume a radial ordering $\pm Im (x-x')>0$ and $|x-x'|\ll L$.
\end{thebibliography}
\end{document}